\documentclass[twocolumn,aps,prl,superscriptaddress,showpacs,byrevtex]{revtex4}
\usepackage{graphicx}
\usepackage{epsfig}
\usepackage{float}
\usepackage{fancybox}
\usepackage{version}
\usepackage{pifont}
\usepackage{amssymb}
\usepackage{bm}

\newcommand{\GeV}{\mbox{GeV}}
\newcommand{\MeV}{\mbox{MeV}}

\begin{document}
				   
%\preprint{\vbox{ %\hbox{BELLE-CONF-0468}
%                 %\hbox{ICHEP04 10-0721}
%                 \hbox{hep-ex/0409005}
%  \hbox{Publication paper draft}
%  \hbox{Version 0524}
%}}

\date{Resubmission version v2.4}

\title{
%\quad\\[0.5cm]
STUDY OF DECAY MECHANISMS IN \bm{${B^-}\to\Lambda_c^+\bar{p}\pi^-$} 
DECAY  AND OBSERVATION OF 
%ANOMALOUS STRUCTURE 
%LOW MASS ENHANCEMENT
LOW MASS  STRUCTURE 
IN THE \bm{$(\Lambda_c^+\bar{p})$} SYSTEM}
%\author{N.\,Gabyshev (Belle collaboration)}
%\address{The Budker Institute of Nuclear Physics,
%Acad. Lavrentiev prospect 11,
%630090, Novosibirsk,
%Russia}

%%% Paper:    B- -> Lambda_c+ pbar pi-
%%% Journal:  Physical Review Letters
%%% Contacts: N. Gabyshev (gabyshev@bmail.kek.jp)
%%%           H. Kichimi (hiromichi.kichimi@kek.jp)
%%%           S. Eidelman (S.I.Eidelman@inp.nsk.su)
%%% Non-responding authors or those who said NO are commented out.
%%% ====================================================================
%%% Click the RELOAD button on your web browser to see the updated file.
%%% ====================================================================
%%% Use \input{author} to insert this material into your latex file.
%%%%% Force institutions to appear in alphabetical order when typeset.
%%%\affiliation{Aomori University, Aomori}
\affiliation{Budker Institute of Nuclear Physics, Novosibirsk}
\affiliation{Chiba University, Chiba}
\affiliation{Chonnam National University, Kwangju}
%%%\affiliation{Chuo University, Tokyo}
\affiliation{University of Cincinnati, Cincinnati, Ohio 45221}
\affiliation{University of Frankfurt, Frankfurt}
\affiliation{Gyeongsang National University, Chinju}
\affiliation{University of Hawaii, Honolulu, Hawaii 96822}
\affiliation{High Energy Accelerator Research Organization (KEK), Tsukuba}
\affiliation{Hiroshima Institute of Technology, Hiroshima}
\affiliation{Institute of High Energy Physics, Chinese Academy of Sciences, Beijing}
\affiliation{Institute of High Energy Physics, Vienna}
\affiliation{Institute for Theoretical and Experimental Physics, Moscow}
\affiliation{J. Stefan Institute, Ljubljana}
\affiliation{Kanagawa University, Yokohama}
\affiliation{Korea University, Seoul}
%%%\affiliation{Kyoto University, Kyoto}
\affiliation{Kyungpook National University, Taegu}
\affiliation{Swiss Federal Institute of Technology of Lausanne, EPFL, Lausanne}
\affiliation{University of Ljubljana, Ljubljana}
\affiliation{University of Maribor, Maribor}
\affiliation{University of Melbourne, Victoria}
\affiliation{Nagoya University, Nagoya}
\affiliation{Nara Women's University, Nara}
\affiliation{National Central University, Chung-li}
%%%\affiliation{National Kaohsiung Normal University, Kaohsiung}
\affiliation{National United University, Miao Li}
\affiliation{Department of Physics, National Taiwan University, Taipei}
\affiliation{H. Niewodniczanski Institute of Nuclear Physics, Krakow}
\affiliation{Nihon Dental College, Niigata}
\affiliation{Niigata University, Niigata}
\affiliation{Osaka City University, Osaka}
\affiliation{Osaka University, Osaka}
\affiliation{Panjab University, Chandigarh}
\affiliation{Peking University, Beijing}
\affiliation{Princeton University, Princeton, New Jersey 08544}
%%%\affiliation{RIKEN BNL Research Center, Upton, New York 11973}
%%%\affiliation{Saga University, Saga}
\affiliation{University of Science and Technology of China, Hefei}
\affiliation{Seoul National University, Seoul}
\affiliation{Sungkyunkwan University, Suwon}
\affiliation{University of Sydney, Sydney NSW}
\affiliation{Tata Institute of Fundamental Research, Bombay}
\affiliation{Toho University, Funabashi}
\affiliation{Tohoku Gakuin University, Tagajo}
\affiliation{Tohoku University, Sendai}
\affiliation{Department of Physics, University of Tokyo, Tokyo}
\affiliation{Tokyo Institute of Technology, Tokyo}
\affiliation{Tokyo Metropolitan University, Tokyo}
\affiliation{Tokyo University of Agriculture and Technology, Tokyo}
%%%\affiliation{Toyama National College of Maritime Technology, Toyama}
\affiliation{University of Tsukuba, Tsukuba}
%%%\affiliation{Utkal University, Bhubaneswer}
\affiliation{Virginia Polytechnic Institute and State University, Blacksburg, Virginia 24061}
\affiliation{Yonsei University, Seoul}
   \author{N.~Gabyshev}\affiliation{Budker Institute of Nuclear Physics, Novosibirsk} % BINP
   \author{K.~Abe}\affiliation{High Energy Accelerator Research Organization (KEK), Tsukuba} % KEK
   \author{K.~Abe}\affiliation{Tohoku Gakuin University, Tagajo} % TohokuGakuin
% \author{N.~Abe}\affiliation{Tokyo Institute of Technology, Tokyo} % TIT
   \author{I.~Adachi}\affiliation{High Energy Accelerator Research Organization (KEK), Tsukuba} % KEK
   \author{H.~Aihara}\affiliation{Department of Physics, University of Tokyo, Tokyo} % Tokyo
% \author{M.~Akatsu}\affiliation{Nagoya University, Nagoya} % Nagoya
   \author{Y.~Asano}\affiliation{University of Tsukuba, Tsukuba} % Tsukuba
% \author{T.~Aso}\affiliation{Toyama National College of Maritime Technology, Toyama} % Toyama
   \author{V.~Aulchenko}\affiliation{Budker Institute of Nuclear Physics, Novosibirsk} % BINP
   \author{T.~Aushev}\affiliation{Institute for Theoretical and Experimental Physics, Moscow} % ITEP
% \author{T.~Aziz}\affiliation{Tata Institute of Fundamental Research, Bombay} % Tata
% \author{S.~Bahinipati}\affiliation{University of Cincinnati, Cincinnati, Ohio 45221} % Cincinnati
   \author{A.~M.~Bakich}\affiliation{University of Sydney, Sydney NSW} % Sydney
% \author{V.~Balagura}\affiliation{Institute for Theoretical and Experimental Physics, Moscow} % ITEP
% \author{Y.~Ban}\affiliation{Peking University, Beijing} % Peking
% \author{S.~Banerjee}\affiliation{Tata Institute of Fundamental Research, Bombay} % Tata
% \author{E.~Barberio}\affiliation{University of Melbourne, Victoria} % Melbourne
% \author{M.~Barbero}\affiliation{University of Hawaii, Honolulu, Hawaii 96822} % Hawaii
% \author{A.~Bay}\affiliation{Swiss Federal Institute of Technology of Lausanne, EPFL, Lausanne} % Lausanne
% \author{I.~Bedny}\affiliation{Budker Institute of Nuclear Physics, Novosibirsk} % BINP
   \author{U.~Bitenc}\affiliation{J. Stefan Institute, Ljubljana} % Ljubljana
% ...
% Ljubljana
% \author{D.~Z\"urcher}\affiliation{Swiss Federal Institute of Technology of Lausanne, EPFL, Lausanne} % Lausanne
\collaboration{The Belle Collaboration}

\begin{abstract}
Using a sample of $152$ million $B\bar{B}$ pairs accumulated with the
Belle detector at the KEKB $e^+e^-$ collider, 
%we perform a Dalitz plot analysis of the
we study the decay mechanism of
three-body charmed decay ${B^-}\to\Lambda_c^+\bar{p}\pi^-$.
The intermediate two-body decay 
${B^-}\to\Sigma_c(2455)^0\bar{p}$ is observed for the first time with a
branching  fraction of 
$(3.7\pm0.7\pm0.4\pm1.0)\times10^{-5}$
and a statistical significance of $8.4\sigma$. 
We also observe a low-mass enhancement in the
$(\Lambda_c^+\bar{p})$ system, which
can be parameterized as a Breit-Wigner function with
a mass of
$(3.35^{+0.01}_{-0.02}\pm0.02)\,\GeV/c^2$ 
and a width of
$(0.07^{+0.04}_{-0.03}\pm0.04)\,\GeV/c^2$. 
We measure its branching  
fraction to be $(3.9^{+0.8}_{-0.7}\pm0.4\pm1.0)\times10^{-5}$ with a
statistical significance of $6.2\sigma$.
  The errors are statistical, systematic,
  and that of the $\Lambda_c^+\to{p}K^-\pi^+$ decay branching fraction.
\end{abstract}

\maketitle

Recently three-body baryon production in charmless $B$ decays
has been studied with the Belle detector
\cite{belle_ppk,belle_Dpp,belle_plambdapi,belle_pph}.
Analysis of these decays shows a common feature: the
invariant mass of the baryon-antibaryon system is peaked near
threshold. 
This feature has generated much theoretical discussion
and may be due to a fragmentation effect, 
production of resonances near threshold or
final state interaction of the produced baryon-antibaryon system
\cite{theory_ppk,theory_Dpp,theory_plambdapi,theory_rosner,theory_kerb}. 
It is of interest to learn whether similar behavior is also observed
in $B$ decays to charmed baryons.
The three-body decay 
${B^-}\to\Lambda_c^+\bar{p}\pi^-$ has been 
previously studied at CLEO~\cite{cleo_blamc} and Belle~\cite{belle_blamc}
with 9.2\,fb$^{-1}$ and  29.1\,fb$^{-1}$ of data, respectively.
%Evidence for $\Sigma_c(2455/2520)^0\bar{p}$ intermediate states was 
%obtained with limited statistics.
%obtained, but no Dalitz plot analysis was
%performed due to the limited statistics.
Here we report analysis of the ${B^-}\to\Lambda_c^+\bar{p}\pi^-$
decay  on a Dalitz plane  based on a data sample of 140\,fb$^{-1}$ 
%corresponding to $(152.0\pm0.7)\times{10}^6$ $B\bar{B}$ pairs,
accumulated at the $\Upsilon(4S)$ resonance with the Belle detector at
the KEKB $e^+e^-$ collider~\cite{KEKB}.

  The large-solid-angle magnetic
  spectrometer Belle described in detail elsewhere~\cite{belle}
  consists of a three-layer silicon vertex 
  detector (SVD), a 50-layer cylindrical drift chamber (CDC),
  a mosaic of aerogel threshold \v{C}erenkov counters (ACC),
  a barrel-like array of time-of-flight scintillation counters (TOF),
  and an array of  CsI(Tl) crystals (ECL) located inside a
  superconducting coil providing a 1.5\,T magnetic field.
  An iron flux return located outside the coil
  is instrumented to detect muons and $K^0_L$ mesons (KLM).
  We use a GEANT based Monte Carlo (MC) simulation to model the
  response of the detector and determine its acceptance\ \cite{sim}.

%pid20060326
 %\color{red}
  The event selection is based on track information
  from the SVD and CDC and particle identification (PID) from the combined
  response of the CDC, ACC and TOF.
  We require the impact parameters of all primary tracks with respect to
  the interaction point (IP) to be within $\pm1$ cm in the transverse $(x-y)$ 
  plane and within $\pm5$ cm in the z-direction (opposite to the $e^+$ beam).
  Proton, kaon and pion candidates are selected using $p/K/\pi$ likelihood
  functions provided by the PID system. We require 
  the likelihood ratios $L_i/(L_i+L_j)$  to be greater 
  than 0.6, where the subscript $i$ denotes the selected particle and 
  $j$ the other two particle species.
  The PID efficiency is 98\% for each track, and the fake probability
  of a pion (kaon) to be identified as a kaon (pion) is less than 5\%. 
  The probability for a pion or kaon to be identified as a proton
  is less than 2\%.
  We detect the $\Lambda_c^+$ via five decay modes: 
  $\Lambda_c^+\to{pK^-\pi^+}$, $p\bar{K^0}$, $\Lambda\pi^+$,  
  $p\bar{K^0}\pi^+\pi^-$ and $\Lambda\pi^+\pi^+\pi^-$.
%\color{black}  
  Inclusion of charge conjugate states is implicit unless otherwise
  stated. 
  Neutral kaons and  $\Lambda$ baryons are reconstructed in the 
  $K^0_S{\to}\pi^+\pi^-$ and  $\Lambda{\to}p\pi^-$
  decay, respectively.
  %Candidate $\Lambda$ baryons are reconstructed in
  %the decay $\Lambda{\to}p\pi^-$.
  The ${B^-}\to\Lambda_c^+\bar{p}\pi^-$ events are identified 
  by their energy difference $\Delta{E}=(\sum E_i)-E_{\rm{beam}}$, and 
  the beam-energy constrained mass 
  $M_{\rm{bc}}=\sqrt{E^2_{\rm{beam}}-(\sum\vec{p}_i)^2}$, where  
  $E_{\rm{beam}}$ is the beam energy, and $\vec{p}_i$ and $E_i$ are the
  three-momenta and energies of the $B$ meson decay products,
  all defined in the $e^+e^-$ center-of-mass system.
  To suppress continuum background, we impose requirements on
  the angle between the thrust axis of the $B$ candidate tracks and
  that of the other tracks and
  on the ratio of the second to the zeroth Fox-Wolfram
  moments\,\cite{fox_wolfram}. 
\begin{figure}[htb]
%\psfrag{(b)}{ }
  \includegraphics[width=0.40\textwidth,bb=0 0 567 280]
{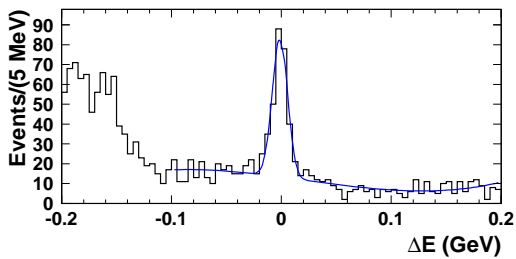}
\vspace{-\bigskipamount}
%\vspace{-\bigskipamount}
\caption{      
  $\Delta{E}$ distribution for $M_{\rm{bc}}>5.27\,\GeV/c^2$
  for ${B^-}\to\Lambda_c^+\bar{p}\pi^-$ candidates.
  The curve shows the result of the fit.
  \label{fig:lamc2_de}
}
\end{figure}

Fig.~\ref{fig:lamc2_de}
shows 
the $\Delta{E}$ distribution for $M_{\rm{bc}}>5.27~\GeV/c^2$ 
for the selected $B$ candidates.
The signal yield is extracted by a fit to the $\Delta{E}$
distribution, as it is free from combinatorial backgrounds from other
$B$ decays. The fit uses a Gaussian for the signal fixed to MC data and a third 
order polynomial for the background for which a MC study shows a broad 
background below the signal due to continuum events and
combinatorial backgrounds from other $B$ decays, such as 
$\overline{B}{}^0\to\Lambda_c^+\bar{p}\pi^0$.
%The $\Delta{E}$ distribution is fitted with
%a Gaussian for the signal fixed to MC plus a third order polynomial
%for the background.
%The background shape is expected
%from a MC study that indicates a broad background
%below the signal due to continuum events and
%combinatorial backgrounds from other $B$ decay
%modes, such as $\overline{B}{}^0\to\Lambda_c^+\bar{p}\pi^0$.
%HK The signal shape parameters are fixed to the values
%HK obtained from a fit to signal MC.
%signal and background components.  
We obtain $264\pm20$ signal events with a statistical
significance of $17.1\,\sigma$.
The uncertainty due to the background parameterization is estimated by
repeating the fit with first and third order polynomials and found to
be small (1.5\%). 
  The significance is defined as
  $\sqrt{-2\ln({\cal{L}}_0/{\cal{L}}_{\rm max})}$, where
  ${\cal{L}}_{\rm max}$
  and ${\cal{L}}_0$ denote the maximum likelihoods with the fitted signal
  yield and the yield fixed at zero, respectively.

\begin{figure}[htb]
%\psfrag{(b)}{(a)}
\includegraphics[width=0.40\textwidth,bb=0 0 567 280]
{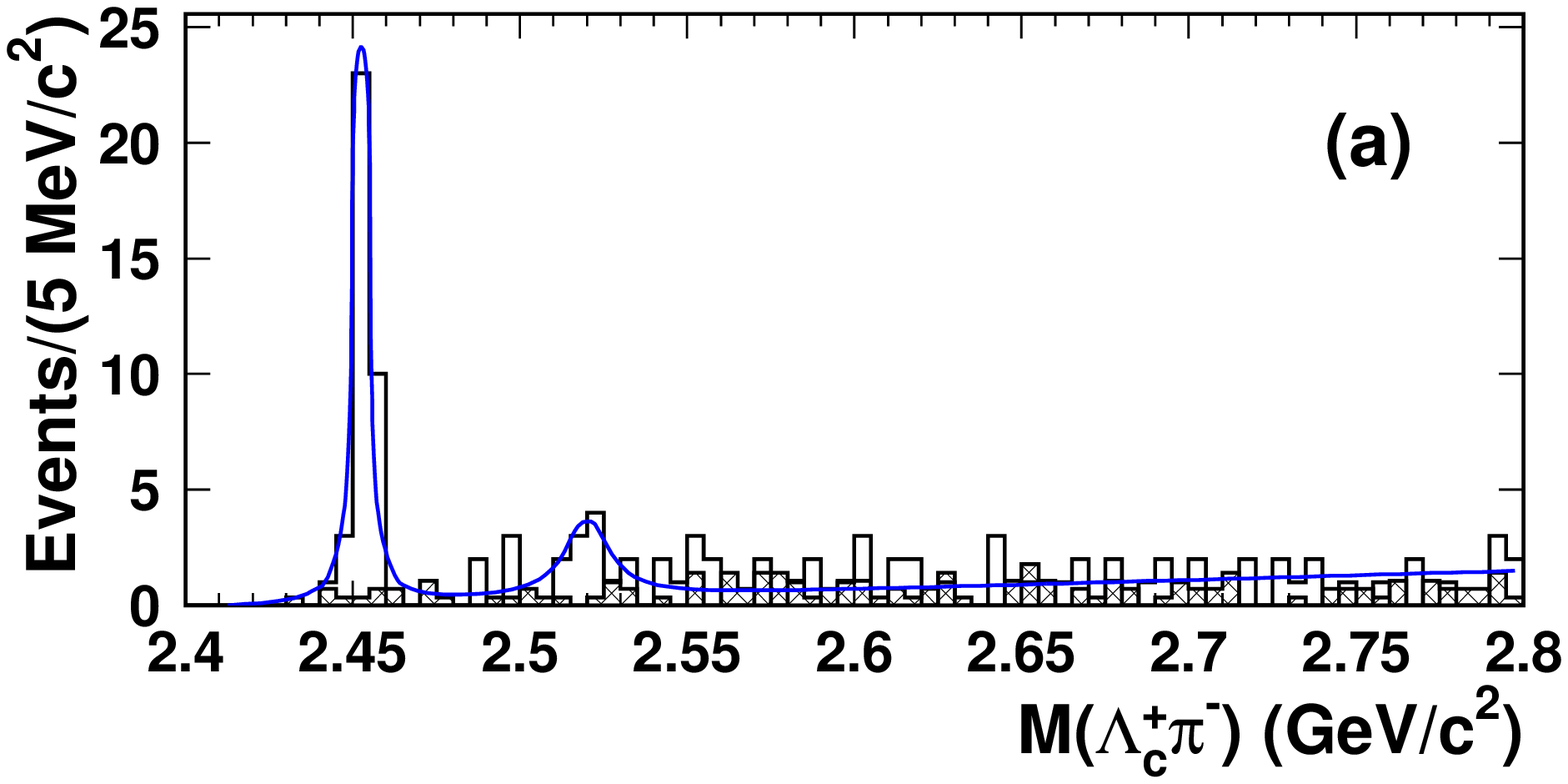} \\
%\psfrag{(c)}{(b)}
\includegraphics[width=0.40\textwidth,bb=0 0 567 280]
{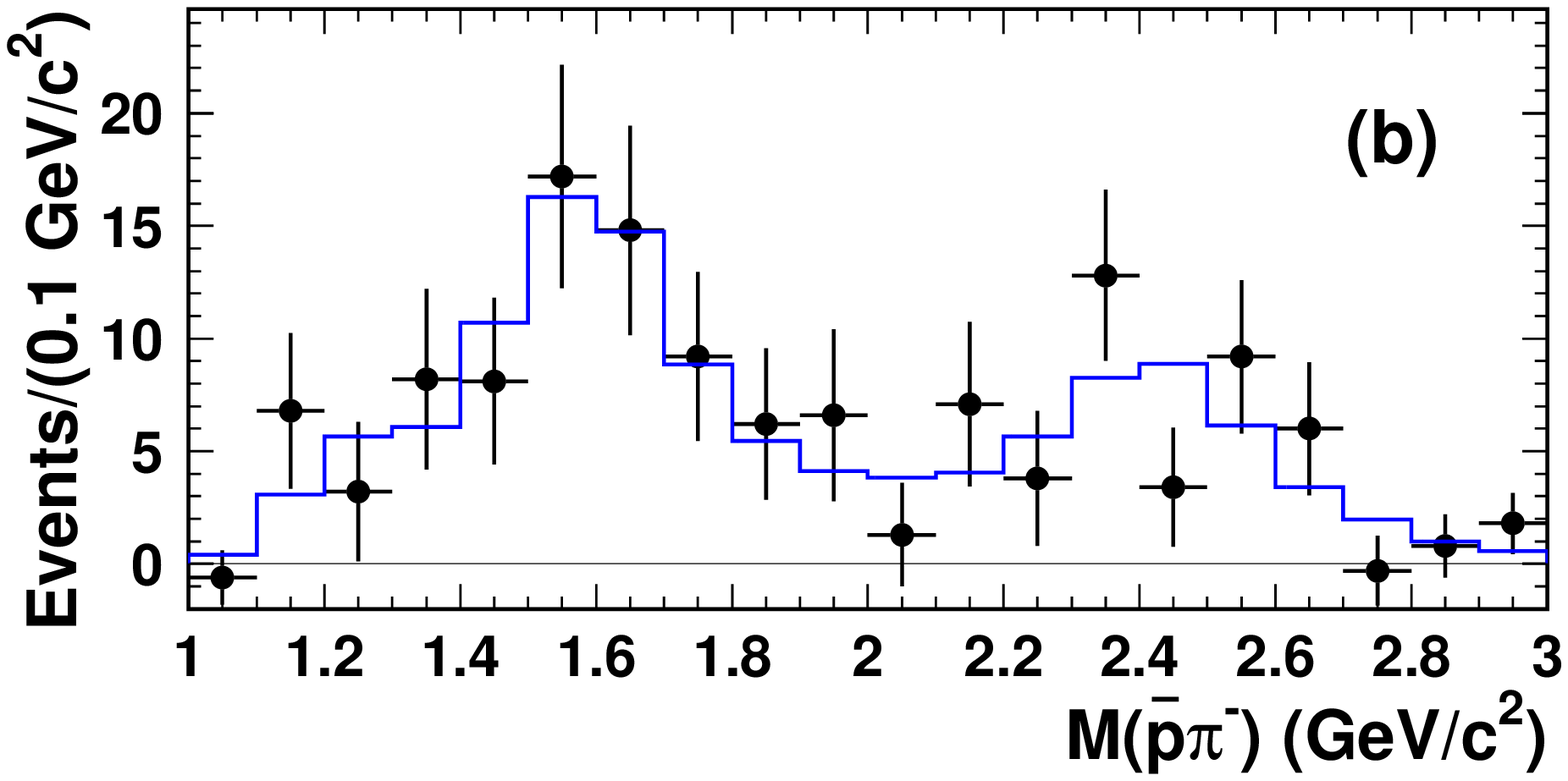} \\
%\psfrag{(d)}{(c)}
\includegraphics[width=0.40\textwidth,bb=0 0 567 280]
{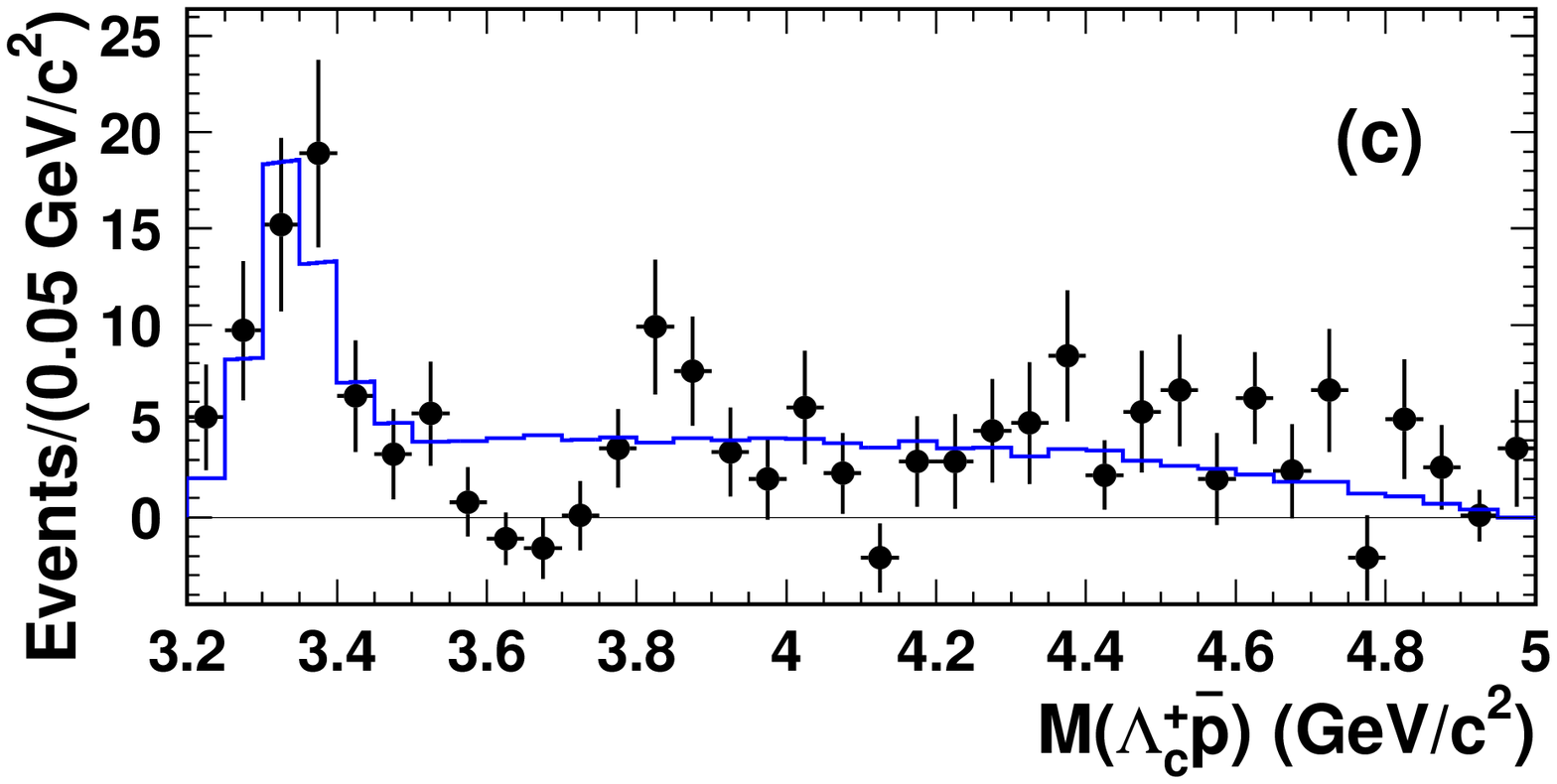}
\vspace{-\bigskipamount}
%\vspace{-\bigskipamount}
\caption{      
  \mbox{(a)} $M(\Lambda_c^+\pi^-)$ distribution for the $B$ signal region
(open histogram) and fit results (curve); the distribution in the 
sidebands is also shown (hatched). 
  \mbox{(b,c)} $B^-$ yields (points) from fits to $\Delta{E}$ distributions,
\mbox{(b)} in bins of $M(\bar{p}\pi^-)$, requiring
  $M(\Lambda_c^+\pi^-)>2.6$~GeV/$c^2$ and 
$M(\Lambda_c^+\bar{p})>3.5$~GeV/$c^2$;
 and \mbox{(c)} in bins of $M(\Lambda_c^+\bar{p})$, requiring
 $M(\Lambda_c^+\pi^-)>2.6$~GeV/$c^2$ and
$M(\bar{p}\pi^-)>1.6$~GeV/$c^2$. The histograms show fit results, 
see the text. 
\label{fig:lamc2}
}
\end{figure}

%Signals from two-body ${B^-}\to\Sigma_c(2455/2520)^0\bar{p}$
%intermediate decays,  evidence for which has  been 
%obtained earlier\,
%\cite{cleo_blamc,belle_blamc},
%are expected in the $M(\Lambda_c^+\pi^-)$  distribution.
%HK Signals from two-body ${B^-}\to\Sigma_c(2455/2520)^0\bar{p}$
%HK intermediate decays are expected in the $M(\Lambda_c^+\pi^-)$  
%HK distribution~\cite{cleo_blamc,belle_blamc}.
Fig.~\ref{fig:lamc2}\mbox{(a)} shows the
$M(\Lambda_c^+\pi^-)$  
%invariant mass 
distribution,
%HK around the $\Sigma_c(2455/2520)^0$ resonances
%HK for ${B^-}\to\Lambda_c^+\bar{p}\pi^-$ decay candidate events
%HK
where clear signals are seen from the intermediate 
two-body ${B^-}\to\Sigma_c(2455/2520)^0\bar{p}$ 
decay~\cite{cleo_blamc,belle_blamc}.
The open histogram is the distribution from the $B$ signal region
({$|\Delta{E}|<0.03\,\GeV$} and {$M_{\rm{bc}}>5.27\,\GeV/c^2$}).
The hatched histogram is the distribution from sideband regions 
({$-0.10\,\GeV<\Delta{E}<-0.04\,\GeV$} or {$0.04\,\GeV<\Delta{E}<0.20\,\GeV$})
normalized to the $B$ signal region area.
The curve shows the result of the fit, which includes the
contribution from $\Sigma_c(2455/2520)^0\to\Lambda_c^+\pi^-$ 
decays and the background parameterized with a linear function.
The $\Sigma_c(2455/2520)^0$ signal shapes are fixed from MC assuming 
a Breit-Wigner function convolved with the resolution function and
using $\Sigma_c(2455/2520)^0$ masses and widths from Ref.~\cite{pdg2002}.
To extract the yields for $B^-{\to}\Sigma_c(2455/2520)^0\bar{p}$ decays,
background from continuum and/or other $B$ decays is taken into account
by fitting simultaneously the $B$ signal and sideband regions in
the $\Delta{E}$ distribution. 
From the fit, we obtain
$35.3^{+6.4}_{-6.0}$ signal events with a statistical significance of
$8.2\,\sigma$ for the $B^-\to\Sigma_c(2455)^0\bar{p}$ decay, and
$12.6^{+5.4}_{-4.7}$ signal events with a statistical significance of
$3.0\,\sigma$ for the $B^-\to\Sigma_c(2520)^0\bar{p}$ decay.

Fig.~\ref{fig:lamc2}\mbox{(b)} shows 
the $M(\bar{p}\pi^-)$ 
%invariant mass 
distribution for the ${B^-}\to\Lambda_c^+\bar{p}\pi^-$ 
candidate events 
from fits to the $\Delta{E}$ distribution in $100\,\MeV$ bins of
$(\bar{p}\pi^-)$ mass
with constraints of
$M(\Lambda_c^+\pi^-)>2.6\,\GeV/c^2$ 
to remove $\Sigma_c^0$ intermediate states
and $M(\Lambda_c^+\bar{p})>3.5\,\GeV/c^2$ 
to remove an enhancement at low $(\Lambda_c^+\bar{p})$,
%invarant mass,
which is discussed below. 
%The histogram is the fit result.
The histogram shows the result of a fit including the following
contributions: 
%We parameterized the distribution with contributions from 
three-body phase space,
${B^-}\to\Lambda_c^+\bar{\Delta}(1232)^{--}$ 
%decay as suggested by theory
\,\cite{lamc2_last_theory}, and two other contributions,
with parameters close to the ${\Delta}(1600)$ and ${\Delta}(2420)$
resonances tentatively referred to as ${\Delta}_{\rm X}(1600)$ and 
${\Delta}_{\rm X}(2420)$. 
The signal shapes are fixed from the MC data.
Both three-body phase space and ${\Delta}(1232)$ contributions 
have a $0.5\sigma$ significance.
The statistical significance of the ${\Delta}_{\rm X}(1600)$ contribution
is $6.7\sigma$ with a yield of $82\pm12$ events while that of 
the ${\Delta}_{\rm X}(2420)$ is $4.7\sigma$ with a
yield of $41\pm9$ events.

Fig.~\ref{fig:lamc2}\mbox{(c)} shows the 
$M(\Lambda_c^+\bar{p})$ 
%invariant mass 
distribution for ${B^-}\to\Lambda_c^+\bar{p}\pi^-$ decay
candidate events 
from fits to the $\Delta{E}$ distribution in $50\,\MeV$ bins of
$(\Lambda_c^+\bar{p})$ mass
with $M(\Lambda_c^+\pi^-)>2.6\,\GeV/c^2$ to remove $\Sigma_c$ contributions
and $M(\bar{p}\pi^-)>1.6\,\GeV/c^2$ to remove low $(\bar{p}\pi^-)$ masses.
A low mass enhancement is observed.
The histogram is the result of a fit parameterizing the 
distribution with a Breit-Wigner peak and feed-downs from
the ${B^-}\to\Lambda_c^+\bar{\Delta}_{\rm X}(1600/2420)^{--}$ contributions.
This fit gives a mass of 
$(3.35^{+0.01}_{-0.02})\,\GeV/c^2$ 
and full width of
$(0.07^{+0.04}_{-0.03})\,\GeV/c^2$. 
The fit yield is $50\pm10$ events with a statistical significance of 
$5.6\sigma$.
%HK
A second peak near $3.8\,\GeV/c^2$ has a 
mass of $(3.84\pm0.01)\,\GeV/c^2$ and width of 
$(0.03\pm0.03)\,\GeV/c^2$.
The yield of this peak is $15\pm6$ events ($2.8\sigma$) and
not studied any further.

Systematic uncertainties are estimated by performing fits with
different background parameterizations
including the contributions of the
${B^-}\to\Lambda_c^+\bar{\Delta}_{\rm X}(1600/2420)^{--}$ feed-downs
with a free number of events or a broad Breit-Wigner function with and without
the second peak at $3.8\,\GeV/c^2$.
The mass variation is less than ${0.02}\,\GeV/c^2$, and
the width varies by less than  ${0.04}\,\GeV/c^2$.

\begin{figure}[htb]
\includegraphics[width=0.40\textwidth,bb=0 0 567 280]
{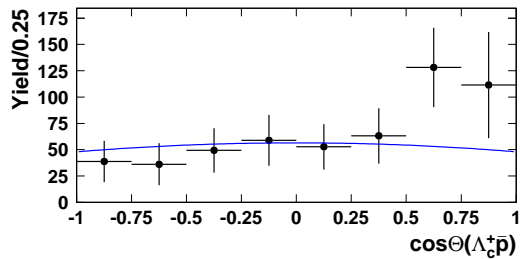}
\vspace{-\medskipamount}
%\vspace{-\bigskipamount}
  \caption{
    The efficiency corrected helicity distribution for the
    $(\Lambda_c^+\bar{p})$ structure. 
    The solid line is the the result of the fit to
    $(1+\alpha\cos^2\Theta(\Lambda_c^+\bar{p}))$. 
%    The dashed line is a fit to a general function 
%    $(1+\alpha\cos^2\Theta(\Lambda_c^+\bar{p}))$ 
%    predicted for the helicity zero state of $(J,J_z)=(1,0)$.
    \label{fig:helicity_exp_lamc2}
  }
\end{figure}

For the $(\Lambda_c^+\bar{p})$ structure, we studied the distribution 
of the helicity angle, $\Theta(\Lambda_c^+\bar{p})$, defined as the
angle between the $\Lambda_c^+$ 
momentum and the direction opposite to the $B$ meson momentum in the 
$(\Lambda_c^+\bar{p})$ rest frame.
%This distribution could distinguish between two possible
%interpretations of the $(\Lambda_c^+\bar{p})$ low mass enhancement.
If  the $(\Lambda_c^+\bar{p})$ structure is due to fragmentation, 
the distribution will be asymmetric, while for a resonance it will be 
symmetric~\cite{theory_rosner} and could provide information on the
spin of the $(\Lambda_c^+\bar{p})$ state.
Fig.~\ref{fig:helicity_exp_lamc2} shows the 
efficiency corrected
helicity distribution for
this decay for events from the region
$M(\Lambda_c^+\bar{p})<3.6~\GeV/c^2$, 
where the data points were
obtained from fits to the $\Delta{E}$ 
distributions.
%The $J=0$ hypothesis is slightly favored (the solid line,
%$\chi^2/\mbox{ndf}=0.97$) 
%over the $J=1$ case (the dashed line, $\chi^2/\mbox{ndf}=1.58$).
The data is consistent with the uniform distribution expected for
a $J=0$ state ($\chi^2/\mbox{ndf}=0.97$). 
A fit of our data to a general formula for the angular
distribution $(1+\alpha\cos^2\Theta(\Lambda_c^+\bar{p}))$ for a $J=1$ 
state~\cite{theory_jpsi_pp,jpsippbar} 
gives $\alpha=(-0.15\pm0.54)$ ($\chi^2/\mbox{ndf}=1.12$),
see the solid curve in Fig.~\ref{fig:helicity_exp_lamc2}.
%For a $J=1$ state, as a helicity zero state 
%is only allowed in the $B\to (\Lambda^+\bar{p})\pi^-$ decay,
%the angular distribution is expressed by
%$(1+\alpha\cos^2\Theta(\Lambda_c^+\bar{p}))$~\cite{jpsippbar}.
%For example, $\alpha=-1$ is for $J/\psi\to\mu^+\mu^-$ in 
%the $B\to J/\psi\pi^-$ decay. 
%The fit (dashed curve) gives $\alpha=(-0.15\pm0.54)$ ($\chi^2/\mbox{ndf}=1.12$).
%As $\alpha$ is unknown 
%for the $\Lambda^+\bar{p}$ system, we do not discuss further.
The observed helicity asymmetry is
$\frac{N_+-N_-}{N_++N_-} = 0.32\pm0.14$, 
where $N_+$ and $N_-$ are the efficiency corrected numbers of events with
$\cos\Theta(\Lambda_c^+\bar{p})>0$ and $<0$, respectively.
No definite conclusions can be drawn from the helicity studies 
with this statistics.

\begin{figure}[htb]
\includegraphics[width=0.40\textwidth]
{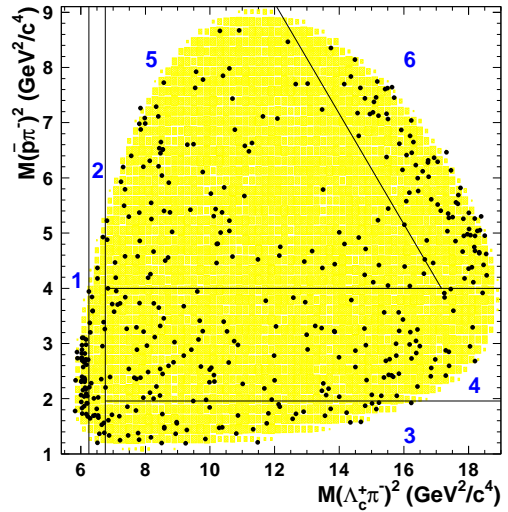}
\vspace{-\medskipamount}
%\vspace{-\bigskipamount}
\caption{
The $M(\bar{p}\pi^-)^2$ vs $M(\Lambda^+_c\pi^-)^2$ Dalitz plot 
subdivided into six regions.
The points are the data in the $B$ signal region.
The shaded area is the phase space MC data.
\label{ref:dal}
}
\end{figure}

We determine the contributions from 
$\Sigma_c(2455/2520)$ and the low ($\Lambda_c^+\bar{p}$) mass
structure, taking into account cross-talk between different resonant
states and the variation of detection efficiency on the Dalitz plane.
%Due to limited statistics,
%we do not consider interference effects among the above contributions except 
%as a possible systematic error, which is treated separately and
%discussed below.
Fig.~\ref{ref:dal} shows
the Dalitz plot of $M(\bar{p}\pi^-)^2$ vs $M(\Lambda^+_c\pi^-)^2$ 
for the events in the $B$ signal region of
{$|\Delta{E}|<0.03\,\GeV$} and {$M_{\rm{bc}}>5.27\,\GeV/c^2$},
where we estimate a background contamination of 37\% in total.
The Dalitz plot is subdivided into six regions corresponding
to the six states discussed above: 
1) $\Sigma_c(2455)^0\bar{p}$ ---
$M(\Lambda_c^+\pi^-)<2.48\,\GeV/c^2$; 
2) $\Sigma_c(2520)^0\bar{p}$ ---
$M(\Lambda_c^+\pi^-)>2.48\,\GeV/c^2$ and
$M(\Lambda_c^+\pi^-)<2.6\,\GeV/c^2$; 
3) $\Lambda_c^+\bar{\Delta}(1232)^{--}$ ---
$M(\Lambda_c^+\pi^-)>2.6\,\GeV/c^2$ and
$M(\bar{p}\pi^-)<1.4\,\GeV/c^2$; 
4) $\Lambda_c^+\bar{\Delta}_{\rm X}(1600)^{--}$ ---
$M(\Lambda_c^+\pi^-)>2.6\,\GeV/c^2$,
$M(\bar{p}\pi^-)>1.4\,\GeV/c^2$ and
$M(\bar{p}\pi^-)<2.0\,\GeV/c^2$;   
5) $\Lambda_c^+\bar{\Delta}_{\rm X}(2420)^{--}$ ---
$M(\Lambda_c^+\pi^-)>2.6\,\GeV/c^2$, 
$M(\bar{p}\pi^-)>2.0\,\GeV/c^2$ and
$M(\Lambda_c^+\bar{p})>3.6\,\GeV/c^2$;
6) $(\Lambda_c^+\bar{p})$ enhancement ---
$M(\bar{p}\pi^-)>2.0\,\GeV/c^2$ and
$M(\Lambda_c^+\bar{p})<3.6\,\GeV/c^2$. 

The resonance parameters are taken from~\cite{pdg2002}. 
%The low mass $(\Lambda_c^+\bar{p})$ enhancement 
The $(\Lambda_c^+\bar{p})$ structure
is represented as a Breit-Wigner function with mass $3.35\,\GeV/c^2$
and full width $0.07\,\GeV/c^2$, 
%as indicated by 
%the above analysis of the $(\Lambda_c^+\bar{p})$.
%invariant mass.
visible as a band in region 6 of Fig.~\ref{ref:dal}.
The $B$ signal yield 
in the $i$-th region is given as 
$X_i = \sum_{j=1}^{6} \varepsilon_{ij} \cdot Y_j$, where
${\varepsilon}_{ij}$ is
the probability to reconstruct the $j$-th intermediate state 
in the $i$-th Dalitz plot region
(estimated from the MC sample of $j$-th state).
We extract the signal $Y_j$ from a simultaneous fit of the $B$ signal
yields $X_{i}$ for the six $\Delta{E}$ distributions, 
where the width is fixed from MC data.
%We extract the signal $Y_j$ from simultaneous fits $B$ meson $\Delta{E}$
%distributions, where the width is fixed from MC data.
Then, we calculate the branching fraction 
${\cal{B}}_j=Y_j/({N_{B\bar{B}}}\times{\eta}\times{\cal{B}}(\Lambda_c^+\rightarrow{pK^-\pi^+}))$
for the $j$-th intermediate state,
combining the $B$ signals tagged with the five $\Lambda_c^+$ decay modes.
$\eta$ is the combined efficiency given by 
$\sum {\eta_k}\times\Gamma_k(\Lambda_c^+)/\Gamma(\Lambda_c^+\rightarrow{pK^-\pi^+})$~\cite{pdg2002}.
$\eta_k$ is the efficiency of the $B$ signal with the $k$-th $\Lambda_c^+$ decay determined from MC.
%The fractions of charged and neutral $B$ mesons are 
%assumed to be equal. 

The resulting yields, branching fractions and statistical significances 
for each intermediate state are listed in
Table~\ref{tab:mfit_res_exp_lamc2}. 
For the branching fractions, the first error is statistical and 
the second is systematic, whereas
the third (26\%) comes from the uncertainty in
${\cal{B}}(\Lambda_c^+{\to}pK^-\pi^+)$.

Systematic uncertainties in the detection efficiencies arise from
the track reconstruction efficiency ($5-7\%$ depending on 
the process, assuming a correlated systematic error of about $1\%$ per 
charged track);
the PID efficiency (about $7\%$ assuming a correlated systematic error of 
$2\%$ per proton and $1\%$ per pion or kaon);
and MC statistics ($1-2\%$).
The other uncertainties are associated with
$\Gamma(\Lambda_c^+)/\Gamma(\Lambda_c^+\to{pK^-\pi^+})$ 
($1-2\%$); 
the number of ${B\bar{B}}$ events ($0.8\%$);
and the parameters of the low mass $(\Lambda_c^+\bar{p})$
enhancement, which can contribute up to $6\%$ to the uncertainty of its
branching fraction.
%(but nothing to the uncertainty of the branching fractions of the
%other intermediate states). 
The total systematic error is estimated to be $9-11\%$ depending on
the intermediate state.

We estimate separately a possible effect of interference between the
different observed intermediate states.
Using special MC samples in which each of
the relative phases among the six states is varied in steps of
$90^\circ$, we compare the signal yield in individual regions of the Dalitz
plot to that obtained using simulated events without any
interference. 
The maximum deviation is treated as the uncertainty
due to interference and is given
in Table~\ref{tab:mfit_res_exp_lamc2}.
%in the last column in Table~\ref{tab:mfit_res_exp_lamc2}.
This simplified treatment of interference does not take into account 
a possibility of reduced compatibility between the simulated
distributions with interference and data and indicates 
that the low mass $(\Lambda_c^+\bar{p})$ enhancement can be partially 
described by such an effect.   

%This simplified treatment of interference indicates 
%that the low mass $(\Lambda_c^+\bar{p})$ enhancement can be partially 
%described by such an effect.   
%The estimate of the effect of interference is conservative and
%does not take into account the possibility of reduced compatibility
%between the simulated distributions with interference and data.

\begin{table*}[htb]
  \caption{
    \label{tab:mfit_res_exp_lamc2}
    Branching fractions for intermediate two-body states in
    $B^-{\to}\Lambda_c^+\bar{p}\pi^{-}$ decay.
%    Detection efficiencies are given for $\Lambda_c^+\to{pK^-\pi^+}$.
  }
\begin{tabular}{lccccc}
\hline
\hline
\multicolumn{1}{c}{Mode} & 
\begin{tabular}{c}
Yield(ev) \\
\end{tabular} &
\begin{tabular}{c}
Significance($\sigma$) \\
\end{tabular} &
\begin{tabular}{c}
Efficiency(\%) \\
\end{tabular} &
\begin{tabular}{c}
Branching($10^{-5}$) \\
\end{tabular} &
\begin{tabular}{c}
Interference error($10^{-5}$) \\
\end{tabular} 
\\
\hline
$B^-{\to}\Sigma_c(2455)^0\bar{p}$ &
$33^{+7}_{-6}$ &
$8.4$ &
$11.7$ &
$3.7\pm{0.7}\pm0.4\pm1.0$ &
$\pm0.6$ \\
$B^-{\to}\Sigma_c(2520)^0\bar{p}$ &
$13^{+6}_{-5}$ &
$2.9$ &
$13.4$ &
$<2.7$ at 90\%CL &
%$1.26^{+0.56}_{-0.49}\pm0.12\pm0.33$ &
$\pm0.8$ \\
%HK &&&&
%HK \\
$(\Lambda_c^+\bar{p})$ structure &
$55^{+11}_{-10}$ &
$6.2$ &
$18.7$ &
$3.9^{+0.8}_{-0.7}\pm0.4\pm1.0$ &
$\pm2.1$ 
\\
$B^-{\to}\Lambda_c^+\bar{\Delta}(1232)^{--}$ &
~$9^{+8}_{-7}$ &
$1.3$ &
$17.9$ &
$<1.9$ at 90\%CL &
%HK $0.65^{+0.56}_{-0.51}\pm0.06\pm0.17$ &
$\pm0.8$ 
%HK \\
%HK &&&&
\\  \hline \hline
$B^-{\to}\Lambda_c^+\bar{\Delta}_{\rm X}(1600)^{--}$ &
%PS1 $\bar{\Delta}_{\rm X}(1600)^{--}$ &
$85^{+15}_{-14}$ &
$7.5$ &
$18.9$ &
$5.9^{+1.0}_{-1.0}\pm0.6\pm1.5$ &
$\pm4.7$ 
\\
$B^-{\to}\Lambda_c^+\bar{\Delta}_{\rm X}(2420)^{--}$ &
%PS2 $\bar{\Delta}_{\rm X}(2420)^{--}$ &
$68^{+15}_{-13}$ &
$6.1$ &
$19.1$ &
$4.7^{+1.0}_{-0.9}\pm0.4\pm1.2$ &
$\pm4.5$ 
\\
\hline \hline
${B^-}\to\Lambda_c^+\bar{p}\pi^-$              & 
$262\pm20$ & $18.1$ &  & $20.1{\pm}1.5{\pm}2.0{\pm}5.2$ \\
\hline
\hline
\end{tabular}
\end{table*}

In summary, using a sample of $152$ million $B\bar{B}$ pairs,
accumulated with the 
Belle detector at the KEKB collider, 
we performed an analysis on the Dalitz plane
of the three-body charmed decay ${B^-}\to\Lambda_c^+\bar{p}\pi^-$.
We report first observation of the two-body
decay mode 
${B^-}\to\Sigma_c(2455)^0\bar{p}$ and measure its branching  
fraction to be $(3.7\pm0.7\pm0.4\pm1.0)\times10^{-5}$
with a statistical significance of $8.4\sigma$.
%The branching fraction for ${B^-}\to\Sigma_c(2455)^0\bar{p}$ is
%comparable to that for the only other known 
%two-body baryonic decay $\bar{B}^0\to\Lambda_c^+\bar{p}$, which was
%discussed in our earlier publication~\cite{belle_lamc1}.
We also observe a low mass enhancement in the
$(\Lambda_c^+\bar{p})$ system,
which can be parameterized as a Breit-Wigner function with
a mass of
$(3.35^{+0.01}_{-0.02}\pm0.02)\,\GeV/c^2$ 
and a width of
$(0.07^{+0.04}_{-0.03}\pm0.04)\,\GeV/c^2$. 
The branching fraction of the $B^-$ decay to this structure
is $(3.9^{+0.8}_{-0.7}\pm0.4\pm1.0)\times10^{-5}$ with a
statistical significance of $6.2\sigma$.
The current data are not sufficient to determine an origin of this
enhancement.
%The current data are not sufficient to determine whether this
%enhancement is a resonance, an effect due to fragmentation
%or a final state interaction of the produced baryon-antibaryon
%system.
%
The total three-body ${B^-}\to\Lambda_c^+\bar{p}\pi^-$ decay branching
fraction has been measured to be  
$(20.1\pm1.5\pm2.0\pm5.2)\times10^{-5}$, which is consistent with
previous results \cite{cleo_blamc,belle_blamc}.
The branching fractions measurements supersede those in
Ref.~\cite{belle_blamc}. 

\begin{acknowledgments}
We thank the KEKB group for the excellent operation of the
accelerator, the KEK cryogenics group for the efficient
operation of the solenoid, and the KEK computer group and
the NII for valuable computing and Super-SINET network
support.  We acknowledge support from MEXT and JSPS (Japan);
ARC and DEST (Australia); NSFC (contract No.~10175071,
China); DST (India); the BK21 program of MOEHRD and the CHEP
SRC program of KOSEF (Korea); KBN (contract No.~2P03B 01324,
Poland); MIST (Russia); MESS (Slovenia); NSC and MOE
(Taiwan); and DOE (USA).
\end{acknowledgments}

  % LocalWords:  DGFs Monte Carlo Wolfram vs picdir CL Det eff Br KEKB
  % LocalWords:  al Eur Phys Masukawa Prog Theor Fu Lett Jarfi Deahpande
  % LocalWords:  Trampetic Chernyak Zhitnitsky Nucl Alimonti Hirano Inst
  % LocalWords:  Akatsu Iijima Kichimi Ikeda Abashian Instr Funakoshi
  % LocalWords:  GEANT lp tex CONF nn Lepton twocolumnfalse Gabyshev
  % LocalWords:  authorlist pc erenkov kinematically misidentication BK
  % LocalWords:  CHEP SRC sloid det Soni DD Meth DPNU Proc Brun Chistov
  % LocalWords:  PACS pK lc lamc dE dx bc iE lr Procario Deshpande
  % LocalWords:  Kikutani bibitem thr Ia IIa IVa blamc eps sim exp
  % LocalWords:  de mb mcprod ECL KLM wideband QQ Ia IIa IIIa CLNS
  % LocalWords:  cccccc lccc lcccc lccccc crrc ccccc lrrc ndf PDFs
  % LocalWords:  mfit minv phsp pbar lc Acad Lavrentiev prospect lll
  % LocalWords:  ij ev Eidelman Chua charmful NII SINET MEXT JSPS DEST
  % LocalWords:  DST NSFC BK MOEHRD KOSEF KBN MIST MESS NSC MOE DOE

\end{document}